\documentclass[fleqn,usenatbib]{mnras}

\usepackage{newtxtext,newtxmath}

\usepackage[T1]{fontenc}

\DeclareRobustCommand{\VAN}[3]{#2}
\let\VANthebibliography\thebibliography
\def\thebibliography{\DeclareRobustCommand{\VAN}[3]{##3}\VANthebibliography}

\usepackage{graphicx}	
\usepackage{amsmath}	

\title[Destroyed dwarfs with the MDF]{Unravelling the mass spectrum of destroyed dwarf galaxies with the metallicity distribution function}

\author[Deason, Koposov et al.]{
Alis J. Deason$^{1,2}$\thanks{E-mail: alis.j.deason@durham.ac.uk (AD)},
 Sergey E. Koposov$^{3,4,5}$\thanks{E-mail: sergey.koposov@ed.ac.uk (SK)},
 Azadeh Fattahi$^{1}$,
 Robert J. J. Grand$^{6,7}$
\\
$^{1}$Institute for Computational Cosmology, Department of Physics, Durham University, South Road, Durham DH1 3LE, UK \\
$^{2}$Centre for Extragalactic Astronomy, Department of Physics, Durham University, South Road, Durham DH1 3LE, UK\\
$^{3}$Institute for Astronomy, University of Edinburgh, Royal Observatory, Blackford Hill, Edinburgh EH9 3HJ, UK \\
$^{4}$Institute of Astronomy, University of Cambridge, Madingley Road, Cambridge CB3 0HA, UK\\
$^{5}$Kavli Institute for Cosmology, University of Cambridge, Madingley Road, Cambridge CB3 0HA, UK\\
$^{6}$Instituto de Astrofisica de Canarias, Calle Via Lactea s/n, E-38205 La Laguna, Tenerife, Spain\\
$^{7}$Departamento de Astrofisica, Universidad de La Laguna, Av. del Astrofisico Francisco Sanchez s/n, E-38206, La Laguna, Tenerife, Spain
}

\date{Accepted XXX. Received YYY; in original form ZZZ}

\pubyear{2023}

\begin{document}
\label{firstpage}
\pagerange{\pageref{firstpage}--\pageref{lastpage}}
\maketitle

\begin{abstract}
Accreted stellar populations are comprised of the remnants of destroyed galaxies, and often dominate the `stellar haloes' of galaxies such as the Milky Way (MW). This ensemble of external contributors is a key indicator of the past assembly history of a galaxy. We introduce a novel statistical method that uses the unbinned metallicity distribution function (MDF) of a stellar population to estimate the mass spectrum of its progenitors. Our model makes use of the well-known mass-metallicity relation of galaxies and assumes Gaussian MDF distributions for individual progenitors: the overall MDF is thus a mixture of MDFs from smaller galaxies. We apply the method to the stellar halo of the MW, as well as the classical MW satellite galaxies. The stellar components of the satellite galaxies have relatively small sample sizes, but we do not find any evidence for accreted populations with $L > L_{\rm host}/100$. We find that the MW stellar halo has $N \sim 1-3$ massive progenitors ($L \gtrsim10^8 L_\odot$) within 10 kpc, and likely several hundred progenitors in total. We also test our method on simulations of MW-mass haloes, and find that our method is able to recover the true accreted population within a factor of 2. Future data sets will provide MDFs with orders of magnitude more stars, and this method could be a powerful technique to quantify the accreted populations down to the ultra-faint dwarf mass scale for both the MW and its satellites.

\end{abstract}

\begin{keywords}
Galaxies: dwarf -- Galaxy: halo -- Local Group -- galaxies: luminosity function
\end{keywords}


\section{Introduction}
Dark matter haloes of all shapes and sizes grow by accumulating lower mass constituents (or subhaloes). The galaxies at the centres of these haloes grow via ongoing star formation, but can also form diffuse `stellar haloes' from the stellar material deposited by the accretion of subhaloes (if they contain stars). Depending on the mass scale, this accreted stellar material can amount to significant (e.g. clusters, $\sim 20-30 \%$) or minuscule (e.g. dwarfs, $\sim 0-5 \%$) fractions of the overall stellar mass of the central galaxy \citep{purcell2007}. Despite having a relatively low stellar mass and surface brightness, stellar haloes retain a record of the lower mass systems that have been digested by haloes over time, and quantifying and understanding this accreted relic has been a major research focus in astronomy for several decades \citep[see e.g.][]{helmi2008,belokurov2013}.

The most-studied stellar halo is, unsurprisingly, that of our own Milky Way (MW) galaxy. However, despite significant progress in recent years, we still only have a qualitative view of the mass spectrum of dwarf galaxies that have been consumed by the MW. Most notably, it has become clear since the game-changing \textit{Gaia} mission \citep{gaia_mission}, that the inner stellar halo (within $\sim 20$ kpc) is dominated by one ancient, massive accretion event, dubbed the \textit{Gaia}-Enceladus-Sausage (GES; \citealt{belokurov2018,helmi2018}). There is also some evidence that an additional massive structure resides in the very central regions of the galaxy (within $\sim 4$ kpc), and was accreted even earlier than the GES \citep{kruijssen2019, horta2021a}. However, it is debated whether or not this is really an accreted structure, or rather in-situ MW material \citep[see e.g.][]{myeong2022,rix2022}. These massive progenitors, join the already discovered streams and substructures, such as the Sagittarius and Orphan streams \citep[e.g.][]{newberg2003, majewski2004,belokurov2007orphan}, and the Virgo \citep{juric2008} and Hercules-Aquila \citep{belokurov2007} clouds (although the latter structures may be related to the GES, see e.g. \citealt{simion2019,chandra2022}), and more stellar structures in the halo are continuously being discovered \citep[e.g.][]{naidu2020}. The overall inventory of the Galactic stellar halo is evolving, but the picture is far from complete, and we have no quantitative `mass-spectrum' of destroyed dwarfs akin to the surviving satellite dwarf luminosity function \citep{koposov2008,tollerud2008,drlica-wagner2020,nadler2020}, which is a pillar of the field.

Many of the halo structures that have been discovered in the MW are identified in phase-space and/or action-angle space. This, of course, is where an astrometric mission such as \textit{Gaia} has enabled a deeper understanding of the phase-space structure of the halo by providing 6D measurements (at least for the inner halo). However, even with perfect 6D data, robustly identifying \textit{distinct} halo substructures is challenging. Indeed, massive progenitors can have several `clumps' in dynamical spaces which cannot be unambiguously disentangled \citep[e.g.][]{callingham2022} and when the stellar material is fully phase-mixed it becomes more difficult to identify from the background \citep[e.g.][]{johnston2008}. Furthermore, even in the space of conserved quantities the clumps may not stay compact due to perturbations from massive systems such as the LMC \citep{koposov2022}. 
This is where chemical information can be crucial, as galaxies of different mass (and star formation history) can have distinct chemical signatures \citep[e.g.][]{venn2004,tolstoy2009}. Most notable, is the well-known mass-metallicity relation for galaxies, which extends down to the dwarf mass scales \citep[e.g.][]{skillman1989, kirby2011}.

More massive galaxies are, on average, more metal-rich, and the relation between mass and metallicity exists over several orders of magnitude in mass \citep[e.g.][]{Tremonti2004, kirby2013}. This relation can, to first order, be explained by the larger gravitational wells of more massive galaxies, which are able to retain metals \citep{dekel1986}. Lower mass galaxies lack the gravity to resist the expulsion of metals due to feedback mechanisms. On the dwarf mass scale, not only does the average metallicity vary with mass, but the width of the metallicity distribution function (MDF) also varies, with the lowest mass dwarfs having a wider spread of metallicities \citep[e.g.][]{kirby2011}. The combined MDF of a population of accreted dwarf galaxies, such as a stellar halo, is therefore the superposition of several individual MDFs. Thus, in principle, metallicity measurements \textit{alone} contain a unique record of the mass spectrum of accreted dwarfs. Indeed, the disentangling of a MDF into its individual components is the main focus of this work. Finally, it is worth noting that previous work on the MDFs of dwarf galaxies has focused on \textit{surviving} dwarfs, which, depending on the largely unknown redshift evolution of the mass-metallicity relation, may or may not be relevant for the destroyed dwarfs that make-up stellar haloes \citep[see e.g.][]{fattahi2020, naidu2022}. 

In this work, we consider Galactic-sized stellar haloes as well as the (potential) stellar haloes of dwarf galaxies. In principle, dwarf galaxies themselves can cannibalize lower-mass dwarfs, and form what we classically think of as a `stellar halo'. However, unlike larger mass scales where the merging dark matter clumps all contain stars, at lower mass scales (below $\sim 10^9M_\odot$ in halo mass) dark matter subhaloes may not have any stars at all \citep[e.g.][]{benitez-llambay2020}. A recent study by \cite{deason2022} showed that the very existence of a stellar halo around a dwarf galaxy can have important implications for both small-scale galaxy formation and the nature of dark matter. For example, the mass-threshold for galaxy formation, which is largely determined by the epoch of reionization, can have a major effect on the stellar haloes of dwarf galaxies: for models with a high mass threshold for galaxy formation ($\gtrsim 10^9M_\odot$) dwarf galaxies should not have stellar haloes at all! Thus, the detection or non-detection of lower mass accretion events surrounding dwarf galaxies, particularly at the ultra-faint mass scale ($M_{\rm star} \lesssim 10^5 M_\odot$), is of utmost importance.

In order to study the MDFs of accreted populations, we need large, ideally unbiased, spectroscopic samples with metallicity measurements. For both the Galactic halo, and dwarf satellite galaxies in the MW, extensive samples are hard to come by, but there has been significant progress in recent years \citep[e.g.][]{kirby2011, lamost, rave_dr5, apogee,walker2007, conroy2019, taibi2022}. Moreover, and importantly, we are entering a new era of spectroscopic surveys in the MW, with several projects such as DESI, WEAVE, 4MOST, and PFS on the horizon \citep{desi_mw, jin2022, 4most, pfs}. Thus, with these new surveys in mind, we develop a new modelling method to extract the mass spectrum of accreted components from a sample of [Fe/H] measurements and apply this to current data sets.

The paper is organized as follows. In Section \ref{sec:model} we outline our methodology and introduce the statistical model. This is a fairly technical section that some readers may want to skip over! The method is applied to spectroscopic samples of classical dwarf satellite galaxies, and Galactic halo data in Section \ref{sec:apps}. We test the method on state-of-the-art cosmological simulations of MW-mass galaxies in Section \ref{sec:sims}, and discuss caveats and future prospects in Section \ref{sec:disc}. Finally, we summarize our main findings in Section \ref{sec:conc}.

\section{MDF modelling}
\label{sec:model}

In this section, we present the methodology that allows us to take samples with measured [Fe/H], and some estimate of the total luminosity of the system, and use them to provide constraints on the number of discrete stellar systems of different luminosities that can contribute to a given galaxy.  

The next section is fairly technical, so a less statistically minded reader may want to skip it and continue with Section \ref{sec:apps}. The {\tt Python} code implementing the inference method presented in this section is released on GitHub\footnote{\url{https://github.com/segasai/mdf_modeling_paper}}.

\subsection{General statistical model}
We construct a generative model that allows us to represent the MDF as a mixture of  MDFs from smaller galaxies.  Throughout this work, we will assume that the MDF of each smaller galaxy can be represented by a Gaussian.

The generic model, where the sample of stars for the MDF is coming from  several galaxies, can be described with these model parameters:
\begin{itemize}
\item Number of galaxies N  
\item $L_i$ individual galaxy luminosities (where $1<i<N$)
\item $\mu_i$ mean galaxy metallicities
\item $\sigma_i$ widths of MDF of individual galaxies.
\end{itemize}

We can then assume that the number of stars in the sample scales linearly with galaxy luminosity. This assumption is accurate for stellar populations of similar ages. For that assumption to hold, our sample must not be biased towards one galaxy or another (e.g. if our sample comes from a small volume that has an unrepresentative subsample of certain galaxies). If the proportionality holds, one can write the MDF as 
\begin{align}
    P(z|N,\{L_i\},\{\mu_i\},\{\sigma_i\}) =\frac{1}{\sum L_i} \sum\limits_{i=1}^{i=N} L_i {\mathcal N}(z|\mu_i,\sigma_i)
    \end{align}

Here, for clarity, we use $z$ as a short-hand notation of [Fe/H]. Given our expectation that galaxy luminosities and metallicities are correlated \citep{Tremonti2004,kirby2011}, we can assume that galaxies follow a mass-metallicity relation (or luminosity-metallicity relation)
\begin{align}
 \mu_i\sim {\mathcal N}(A +B \log L_i |{\mathcal S})
\end{align}

where $A$ and $B$  are constants i.e. taken from the mass-metallicity relation presented in  \cite{kirby2011} and \cite{simon2019}. $\mathcal S$ is a constant representing a scatter in the relation \citep[found to be 0.15 dex by][for MW satellites]{simon2019}.

The individual widths $\sigma_i$ of MDFs differ from galaxy to galaxy but have been approximated to be slowly dependent on the galaxy luminosity 
$\sigma = C+D \log L$ \citep[see][]{simon2019}. If we specify the constants $A$, $B$, $C$, $D$, and $\mathcal S$ we have a model for the distribution of metallicities, and this model has an integer parameter $N$ and $2 N$ floating point parameters for luminosities and metallicities of $N$ individual galaxies.

While this model for the MDF is valid and can be applied to real data, it has the problem of having a variable number of parameters and therefore is  difficult to sample in practice \citep[i.e.][]{green1995}. Therefore, it would be beneficial to reformulate the model in a way that makes the number of parameters fixed. 

The first modification we can do is to group galaxies in $M$ luminosity bins, so that rather than representing their luminosities by discrete parameters we represent the number of galaxies in certain luminosity bins. Now we define:

\begin{itemize}
    \item 
${\hat L}_j$ are the grid of galaxy luminosities $1\le j \le M$
\item $N_j$ are the numbers of galaxies with luminosities ${\hat L}_j$.
\item $\mu_{j,k}$ are mean metallicities of k-th galaxy with luminosity ${\hat L}_j$. $1<k<N_j$
\end{itemize}

Where due to mass-metallicity relation $$ \mu_{j,k}\sim {\mathcal N}(A+B \log {\hat L}_j |\mathcal S) $$
or $$\mu_{j,k} = A+B \log {\hat L}_j  + {\mathcal S} \epsilon_{j,k}$$
where $\epsilon_{j,k}\sim {\mathcal N}(0,1)$. Here ${\mathcal S}$ could either be a constant or a deterministic function of ${\hat L}_j$

The MDF model is now 
$$P\left(z|\{N_j\},\{\epsilon_{j,k}\}\right) =\frac{1}{\sum{N_j L_j}} \sum\limits_{j=1}^{j=M} {
\hat L}_j \left [\sum\limits_{k=1}^{k=N_j} {\mathcal N}(z|\mu_{j,k},\sigma_{j,k})\right]$$.

The likelihood of the data consisting of (for simplicity) a single star with metallicity z would be exactly $P(z|\{N_j\},\{\epsilon_{j,k}\})$. 
The only problem with this formulation is that this likelihood still depends on a variable number of parameters $\epsilon_{j,k}$ so one would prefer to marginalize over these.

$$P(z|\{N_j\}) =\int  P(z|\{N_j\},\{\epsilon_{j,k}\}) {\mathcal  N}(\{\epsilon_{j,k}\}|0,1)  d \epsilon_{j,k}$$

While this marginalization is difficult, and may be impossible to do analytically, one can simply perform a Monte-Carlo integration over $Q$ samples from a normal distribution, where 
$\epsilon_{j,k,q}$ are the q-th sample $1\le q\le Q$ from ${\mathcal N}(0,1)$

$$P(z|\{N_j\}) \approx \frac{1}{Q} \sum\limits_{q=1}^{q=Q} P(z|\{N_j\},\{\epsilon_{j,k,q}\}) $$

Finally, instead of directly doing the summation we can simply treat this as likelihood with integer parameter $q$

\begin{align}
P(z|\{N_j\}, q) =  P(z|\{N_j\},\{\epsilon_{j,k,q}\}) 
\label{eqn:like}
\end{align}

where $q$ is a nuisance seed parameter that we marginalize over under the uniform prior $U(1,Q)$. Here, we assume that $\epsilon_{j,k,q}$ are coming from a pseudo-random number generator that is seeded by $q$ and provides normally distributed samples.  We then will need to sample the posterior over $\{N_j\}$ and $q$, which gives the model with $M+1$ parameters. Armed with Eqn~\ref{eqn:like} that specifies the likelihood function for the metallicity distribution, the only missing ingredient for the model are the priors.

We assume that occupation numbers $\{N_i\}$ (i.e. numbers of galaxies in luminosity bins) have a prior distribution of  $\lfloor 10^x\rfloor$ where $x\sim U(-1,4)$. This is essentially the log uniform of integers  distribution with 20\% prior volume at $N_i=0$  and 20\% for $1\le N_i\le 10$, and $10\le N_i \le 100$ etc.

Finally, we complement the model with the constraint on the total luminosity of the system. Specifically, we require that the combined luminosity of multiple galaxies must match certain known  total luminosity $\log L_{\rm tot} $ with some uncertainty $\sigma_L$. 
This provides a term for the log of the posterior. 

$$\log (\sum N_i \hat{L}_i) \sim  N(\log L_{\rm tot}|\sigma_L)$$

A final remark that despite the introduction of the formalism based on binned number of galaxies, we have found the model is more stable when at least one contributor to the MDF (likely the one being the most massive main progenitor) is represented directly (rather than in a bin) by the satellite luminosity $L_{\rm main}$, metallicity $z_{\rm main}$ and that also obeys the mass-metallicity relation.

\begin{figure}
    \centering
    \includegraphics{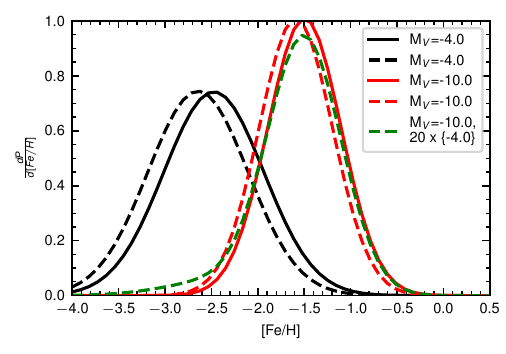}
    \caption{The simulated MDFs for a few systems of different luminosities. The black lines show the expected MDFs in our model for a system with $M_V=-4$, with solid and dashed curves showing the MDFs when using a different random seed that controls the offset of the galaxy with respect to the mass-metallicity relation.  Red curves similarly show the MDF of a single $M_V=-10$ galaxy with different random seeds. The green curve shows the MDF for a synthetic galaxy that consists of stars coming from one galaxy with $M_V=-10$ and 20 galaxies with $M_V=-4$. }
    \label{fig:mdf_model}
\end{figure}

To illustrate our modelling approach, in Figure~\ref{fig:mdf_model} we show the expected [Fe/H] distributions given our model. Specifically, solid black and red curves show possible MDFs for a single galaxy of $M_V=-4$ and $M_V=-10$, respectively. Dashed lines of the same colours show the MDFs when different random seeds are used. The green curve shows a distribution that we might expect if we observe stars coming from a single $M_V=-10$ galaxy and 20 $M_V=-4$ systems. 
This shows a prominent tail towards low metallicities, and this is exactly what allows us to probe the number of possible mergers with low luminosity systems.

\subsection{Sampling}

In the previous section, we have introduced the likelihood function for the metallicity distribution that is conditional on the number of different dwarf galaxies ${N_j}$ on a grid of luminosities. The model also has an integer seed parameter $q$. 
It is not trivial to sample integer parameters, especially if we expect multiple modes. 
To perform the sampling we decide to use the dynamic nested sampling as implemented in the {\tt dynesty} package \citep{speagle2020,dynesty123}. As nested sampling is technically invalid if the likelihood surface has plateaus \citep{fowlie2020}, we add a small level of deterministic noise with standard deviation of 0.01 to the likelihoods, which should not affect the inference.\footnote{The recently released 2.1.0 version of {\tt dynesty} lifts the limitation and is now able to sample likelihood functions with plateaus.}  

\begin{figure*}
    \centering
    \includegraphics{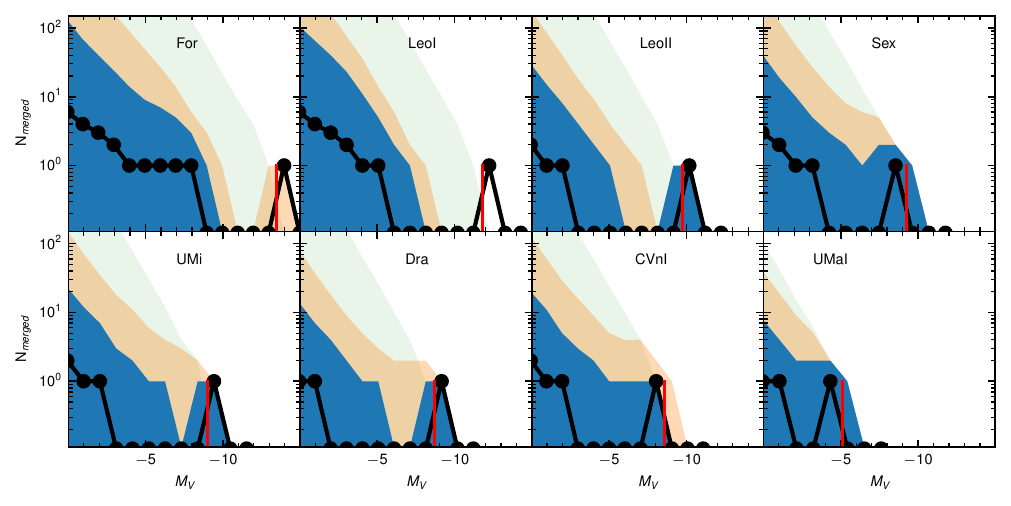}
    \caption{The inferred contributions from systems to the MDF of different dwarf galaxies from our analysis. In each panel, the black curve shows the median number of galaxies of a given luminosity that could have contributed to the MDF. The blue and orange bands show the 16/84 and 1/99 percentiles, respectively. The green band shows the sampling of the prior with only the constraint on total luminosity of the system. The vertical red line on each panel shows the luminosity of each system. Note that the logarithmic y-axis is cut-off at $N_{\rm merged}=10^{-1}$, so median values at this level are consistent with zero.}
    \label{fig:mw_dwarfs_diff}
\end{figure*}

\begin{figure*}
    \centering
    \includegraphics{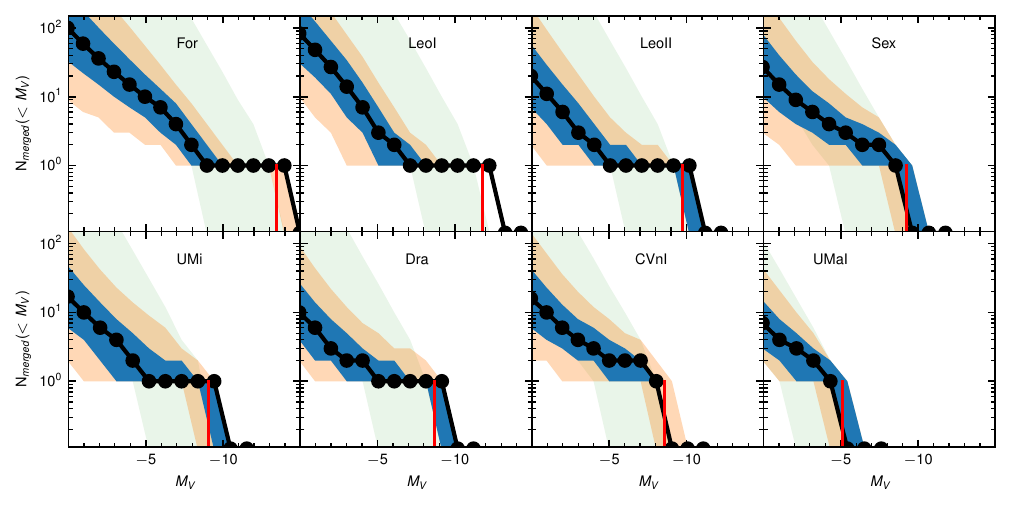}
    \caption{The inferred contributions to the MDF from our analysis. This is similar to Figure~\ref{fig:mw_dwarfs_diff} but shows the cumulative numbers. Each panel shows a different dwarf galaxy. In each panel, the black curve shows the median number of galaxies of a given luminosity or brighter that could have contributed to the MDF. The blue and orange bands show the 16/84 and 1/99 percentiles, respectively. The green band shows the sampling of the prior with only the constraint on total luminosity of the system. The vertical red line on each panel shows the luminosity of each system. }
    \label{fig:mw_dwars_cumul}
\end{figure*}

\section{Applications}
\label{sec:apps}

We now apply the method described above to observational data. Here, we focus on the classical MW satellites (\S\ref{sec:mw_sats}) and the MW stellar halo (\S\ref{sec:mw_halo}).

\subsection{Classical dwarf satellite galaxies}
\label{sec:mw_sats}
We start from the homogeneous sample of dwarf galaxy members presented in \citet{kirby2011} as provided in the Strasbourg astronomical Data Center (CDS). As mentioned in the previous section, the key assumption that we rely on for our method is that the abundances that we model are random samples from the system. This is likely not technically correct for the data at hand since the stellar samples in dwarfs tend to be biased towards the centres of systems \citep[see e.g.][]{walker2011}, and may have slight metallicity biases caused by the colour-magnitude selection of spectroscopic targets. We will, however, proceed ignoring these issues. 

We take the sample of stars from \cite{kirby2011} and only consider stars with  small metallicity uncertainty $\sigma_{[\mathrm{Fe/H]}}<0.2$. This catalogue has measurements of 10 MW satellites with more than 10 stars: Canes Venatici I, Draco,  Fornax, Hercules, Leo I, Leo II, Sculptor, Sextans, Ursa Minor and Ursa Major I. We then proceed to model each of the dwarfs with the machinery presented in Section \ref{sec:model}.  We take the luminosities of each system from \cite{mcconnachie2012} (using an updated catalogue from January 2021) and adopt an $M_V$ uncertainty for each system of 0.1 mag.  For each system, we use the luminosity bins that are 1 mag wide from $M_V=0$ to the luminosity of the dwarf itself minus 2.5 mag.  

The posterior samples on the number of possible dwarf galaxies that contributed to the systems' MDF are shown  in Figures~\ref{fig:mw_dwarfs_diff} and \ref{fig:mw_dwars_cumul}.
We show measurements for 8 out of 10 systems spanning the luminosity range from $M_V\sim-5$ for Ursa Major I to $M_V\sim -13$ for Fornax. The panels are ordered by system luminosity. The total number of stars varies from $N=15$ for Ursa Major I to $N=789$ for Leo I. 
Figure \ref{fig:mw_dwarfs_diff} shows the constraints on the differential number of systems that have contributed to the dwarfs' MDF, while Figure \ref{fig:mw_dwars_cumul} shows constraints on the cumulative counts of the number of systems brighter than a certain value. The blue/orange bands show the 16/84 and 1/99 percentiles, and the black line shows the median of the posterior. The green bands show the constraints if we do not use metallicities at all. This is essentially a prior and corresponds to the case where the only constraint comes from ensuring the combination of galaxies matches the total luminosity of the system.  Note that, because we include all the stars in the galaxy, we expect to measure $N_{\rm merged} = 1$ at around the total luminosity of the dwarf galaxy (shown with the solid red line). Although technically this is an `in-situ' rather than an accreted component, what we are actually constraining are the contributors to the MDF, regardless of their origin.

We now look at the posteriors in more detail. First, we focus on clear cases where the data is particularly constraining. These are the cases of Fornax, Leo I, Leo II, and Draco, where the spectroscopic samples have hundreds of members. We see that their differential posterior distributions (Figure~\ref{fig:mw_dwarfs_diff}) have a peak with a value of one next to the system luminosity (highlighted in red) and show the value consistent with zero for $M_{V, \rm host} \lesssim M_V\lesssim M_{V,\rm host}+5$. Thus, the data suggests that these systems did not experience a merger with a dwarf that is larger than 1\% of the system luminosity. This is also seen in the cumulative plots, where we see the implied number $N_{\rm merged}(<M_V)$ is flat and equal to one in the range $M_{V,\rm host} \lesssim M_V\lesssim M_{V,\rm host}+5$. Looking at the implications for the number of faint contributors to the MDF for Fornax, Leo I, Leo II, and Draco systems, we can see that our constraints on $N_{\rm merged}$ shoot up and become significantly broader. The differential counts are essentially unconstrained. For example, for the Fornax MDF contributors at $M_V=0$ (top left-hand panel of Figure~\ref{fig:mw_dwarfs_diff}) the 1-$\sigma$ confidence interval is $0<N_{\rm merged} <100$ as the data allows many faint dwarfs before the observed MDF is affected significantly. The behaviour of the cumulative counts for the faint MDF contributors is somewhat misleading as it rises at faint $M_V$ purely because we are summing over bins with non-negative values.

Fainter dwarf galaxies like CVnI or UMa have a smaller number of spectroscopic observations. In Figure \ref{fig:mw_dwarfs_diff} we see that the posteriors on the number of MDF contributors start to rise next to $M_V=M_{V,\rm host}$, which indicates that we cannot even rule out that the galaxy is a product of a merger of two systems with similar luminosities.  The constraints on the cumulative number of mergers for fainter dwarfs  do not show a flat $N_{\rm merged}=1$ part next to the luminosity of the system and instead rises to faint luminosities. We also see that for faint systems, the posteriors basically look very close to priors.

\subsection{Galactic stellar halo}
\label{sec:mw_halo}
\begin{figure}
    \centering
    \includegraphics[width=\linewidth,angle=0]{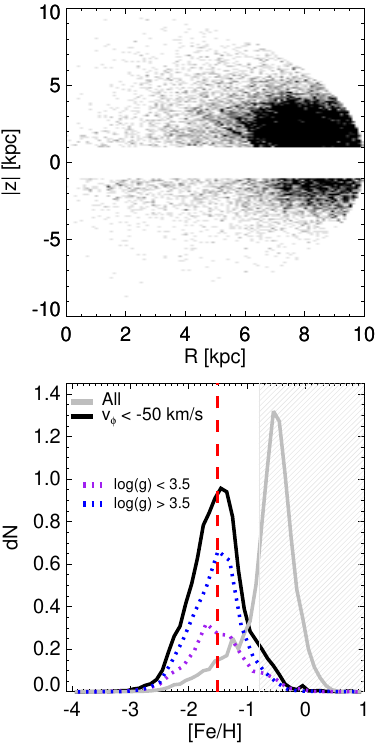}
    \caption[]{\textit{Top panel:} The spatial distribution in the $z$ vs. $R$ plane of our MW halo sample. \textit{Bottom panel:} The metallicity distribution of the sample. The grey line indicates the MDF without a cut in $v_\phi$, which leads to the inclusion of (metal-rich) disk stars. The dashed red line indicates the median metallicity of our halo sample ([Fe/H] $=-1.5$). We also show the MDFs of our halo sample split by log$(g)$ with the blue and purple dotted lines, respectively. The gray line-filled region indicates the metal-rich regime ([Fe/H] $>-0.8$) that is excluded in our modelling.}
    \label{fig:mw_sample}
\end{figure}

\begin{figure*}
\begin{minipage}{\linewidth}
    \centering
    \includegraphics[width=\textwidth,angle=0]{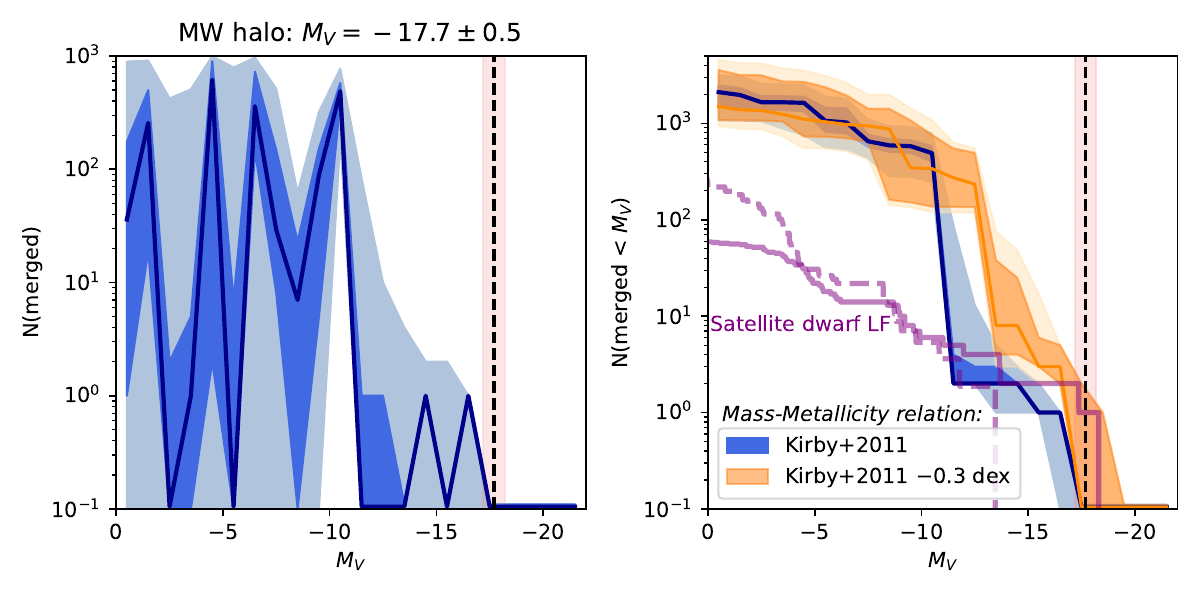}
    \end{minipage}
  \begin{minipage}{\linewidth}
  \centering
   \includegraphics[width=\textwidth,angle=0]{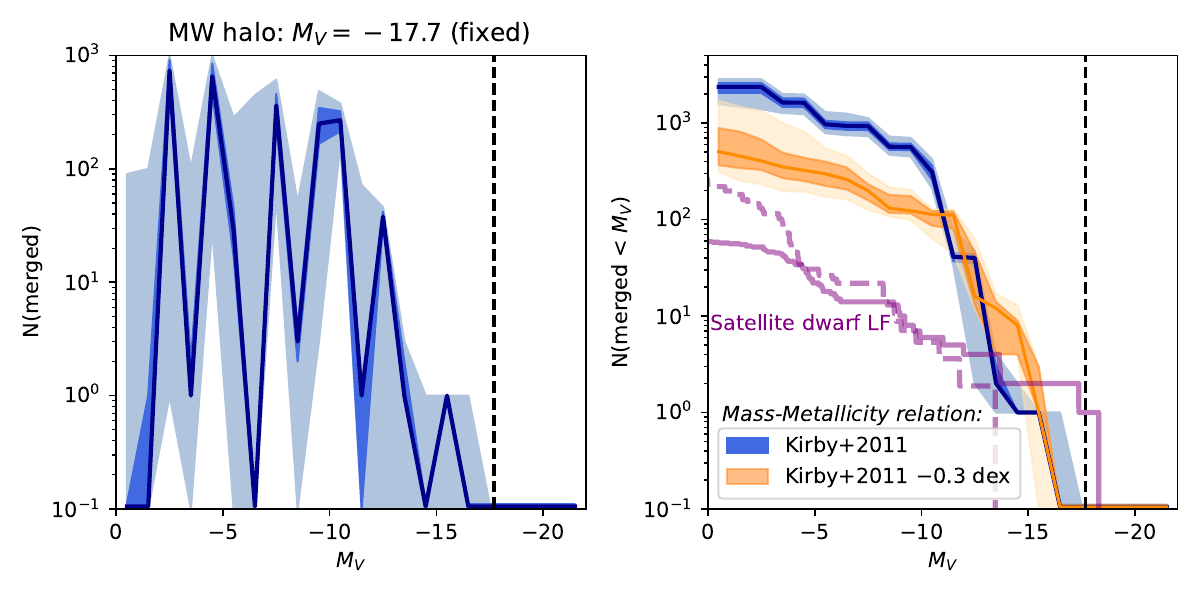}
\end{minipage}
    \caption[]{The estimated differential (left) and cumulative (right) number of destroyed dwarfs in the MW halo. The dark(light) shaded regions show the 16-84(1-99) percentiles, and the solid lines are the medians. The dashed black line indicates the assumed total stellar halo luminosity ($M_{V} = -17.7$). In the top panels, the total luminosity has a flexible uncertainty of $\pm 0.5$ dex, whereas in the bottom panel the total luminosity is kept fixed. The results in blue are for when the $z=0$ mass-metallicity relation for dwarfs is assumed \citep{kirby2011}. In orange, we show the results when an $-0.3$ dex offset is applied to the relation, which has been postulated to be more applicable to destroyed dwarfs \citep{naidu2022}. For comparison, we show the surviving dwarf satellite luminosity function in purple. The dashed line indicates the completeness-corrected LF derived by \cite{drlica-wagner2020}.}
    \label{fig:mw_halo}
\end{figure*}

\begin{figure}
    \centering
    \includegraphics[width=\linewidth,angle=0]{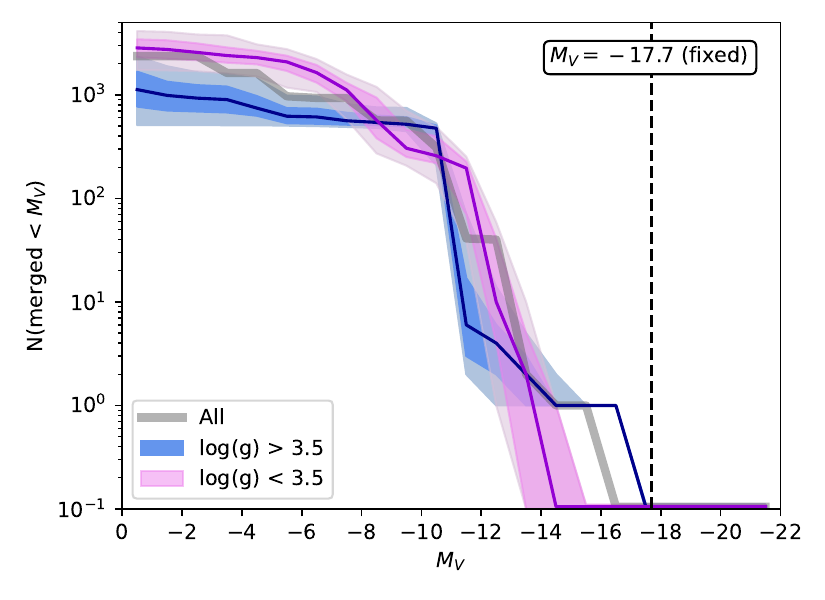}
    \caption[]{The estimated cumulative number of destroyed dwarfs in the MW halo. Same as Fig. \ref{fig:mw_halo}, but split into two bins with low (log(g) $<3.5$) and high (log(g) $>3.5$) surface gravity stars. The thick gray line indicates the overall sample. The stellar halo luminosity is fixed ($M_V=-17.7$).}
    \label{fig:mw_halo_logg}
\end{figure}

We next apply our method to the Galactic stellar halo. It has been realized for some time that the stellar halo of the MW comprises an assortment of destroyed dwarf debris, and thus the metallicity distribution of these halo stars retains a memory of their dwarf galaxy progenitors. 

Large, homogeneous samples of halo stars with metallicity measurements are hard to come by, and this is a significant limitation of our current study. At present, we build a sample of halo stars based on several spectroscopic surveys and use the latest \textit{Gaia} data \citep{gaia_edr3}, to help select a clean halo sample. We begin by cross-matching stars with spectroscopic data from SDSS \citep{sdss_dr14}, RAVE \citep{rave_dr5}, LAMOST \citep{lamost}, APOGEE \citep{apogee}, and GALAH \citep{galah_dr3} with \textit{Gaia} DR3. This results in $N=6,560,819$ stars. To estimate distances to the stars we use the \cite{bailer-jones2021} photogeometric distances computed from \textit{Gaia} EDR3. We only consider stars with reasonable parallax $\sigma_{\varpi} / \varpi <0.5$, and restrict our sample to $r < 10$ kpc and $|z|$ > 1 kpc. Finally, to avoid disk contamination, we apply a cut on the rotational velocity of the stars. We impose a fairly strict cut to remove the majority of thick disk and/or \textit{splash} stars \citep{belokurov2020}, and only include those with retrograde orbits $v_\phi < -50$ km/s. The resulting spatial (top-panel) and metallicity distribution (bottom-panel) of the stars are shown in Fig. \ref{fig:mw_sample}. In the bottom panel, we also show the MDF for the stars without the $v_\phi$ cut in grey. Our restriction to retrograde orbits is fairly stringent but, as can be seen in the figure, it is effective at removing disk stars, which have prograde orbits and are generally more metal-rich. We apply our modelling procedure to stars with $-4<$ [Fe/H] $<-0.8$\footnote{Note that this metallicity cut is made in both the data \textit{and} model, so there is no metallicity bias introduced with our selection.}, and $\sigma$([Fe/H]) $<0.2$, which results in a sample of $N=21,813$ stars.

Our sample is comprised of 5 different spectroscopic surveys, with varying selection functions. Here, we aim to maximise the number of halo stars with metallicity measurements by combining these surveys but note that, ideally, a more homogeneous sample would be used. For now, we continue on, under the assumption that there are no significant metallicity biases in this combined sample. However, we stress that future work with upcoming spectroscopic surveys such as DESI \citep{desi_mw} and WEAVE \citep{weave} will be much better suited for this type of analysis.

In our analysis, we adopt a total halo luminosity of $M_V=-17.7 \pm 0.5$. This is consistent with recent measurements which suggest $L \sim 1 \times 10^9 L_\odot$, and also allows a range around this value encompassing the majority of observational constraints and their uncertainties \citep{deason2019, mackereth2020, horta2021}. The top panels of Fig. \ref{fig:mw_halo} show the resulting number of destroyed dwarfs in the MW halo as a function of $M_V$. We consider dwarfs with $0 > M_V > -22$, using 22 bins with 1 mag bin size. Note that the size of our sample means that we are unlikely constraining dwarfs with $M_V \gtrsim -8$, which will not be represented by a large enough number of stars (see e.g. Section \ref{sec:future}). In blue, we show the results when the \cite{kirby2011} mass-metallicity relation is used, which is appropriate for surviving dwarf galaxies in the MW. In recent work, \cite{naidu2022} (see also \citealt{fattahi2020}) argue that \textit{destroyed} dwarfs may not lie on this relation, and a relation with $\sim 0.3$ dex offset to lower metallicities is more appropriate. We show the results with this offset applied in orange. 

Our model predicts several hundred ($N \sim 400$) destroyed dwarfs with $M_V \lesssim -10$. However, the different mass-metallicity relations (relevant for either `surviving' or `destroyed' dwarfs) predict different distributions of progenitor masses, particularly at larger masses. For example, when using the \cite{kirby2011} mass-metallicity relation applicable for surviving dwarf galaxies, we estimate $N = 1$ massive dwarf progenitor with $L \sim 10^{8.5}L_\odot$, but this rises to $N= 3$ when the relation more relevant to destroyed dwarf galaxies is used instead.  This seems to be at odds with our adopted \textit{total} halo luminosity of $L \sim 1 \times 10^9L_\odot$. Indeed, by summing the predicted numbers of destroyed dwarfs we find that the total luminosity when the \cite{kirby2011} relation is used is $1.1^{+0.2}_{-0.2}  \times 10^9 L_\odot$, but this rises to $3.4_{-2.3}^{+7.2} \times 10^9L_\odot$ when an $0.3$ dex offset is applied to the mass-metallicity relation. Clearly, in this latter case, the bias in metallicity has pushed the progenitor masses higher, and, because we have allowed a fairly flexible total luminosity, resulted in a high halo luminosity. However, it is still consistent with the input luminosity within $1-\sigma$. 

In the bottom panel of Fig. \ref{fig:mw_halo} we show the results when the total halo luminosity is fixed to $M_V=-17.7$ (technically, an uncertainty of $0.01$ dex is adopted). Here, the `fiducial' result using the \cite{kirby2011} mass-metallicity relation is only slightly changed. For example, the most massive progenitor is shifted to a slightly lower luminosity (by $\sim 1$ dex in $M_{V}$), and the total number of dwarfs with $M_V < -10$ is reduced ($N \sim 300$). 
In general, the changes are within the predicted uncertainties. When an 0.3 dex metallicity offset is applied to the mass-metallicity relation, fixing the halo luminosity has a greater effect. This is unsurprising given that allowing for a more flexible halo luminosity favours a higher value than the fiducial $1 \times 10^{9}L_\odot$. In this case, the most massive progenitor has $L \sim 10^{8.1}L_\odot$ (compared to $N\sim 3$ with $L \sim 10^{8.5}L_\odot$ when the total luminosity is more flexible). The number of low-mass dwarfs is also reduced, with $N \sim 110$ with $M_V < -10$.  This exercise emphasizes how important the assumed total halo luminosity, as well as the adopted mass-metallicity relation are for this type of analysis.

We also show the surviving dwarf satellite luminosity function for comparison in Fig. \ref{fig:mw_halo}. Here, we show the observed (solid purple) and completeness-corrected (dashed purple) cumulative number counts given by \cite{drlica-wagner2020}. Note that the completeness-corrected counts differ from the `observed' counts at the bright end because it does not include the massive satellites (Sagittarius, SMC and LMC). The numbers of low luminosity satellite systems are much lower than the predicted number of destroyed dwarfs. This is perhaps unsurprising given that our estimates are likely overestimated at low luminosities, owing both to sample size and our assumption of Gaussian MDFs (see Section \ref{sec:caveats}). Interestingly, the (cumulative) number counts are similar at intermediate luminosities ($-16 \lesssim M_V \lesssim -12$) but destroyed dwarfs as massive as the LMC are not favoured unless the adopted mass-metallicity relation is adjusted from the fiducial $z=0$ form.

\begin{figure*}
    \centering
    \includegraphics[width=\textwidth,angle=0]{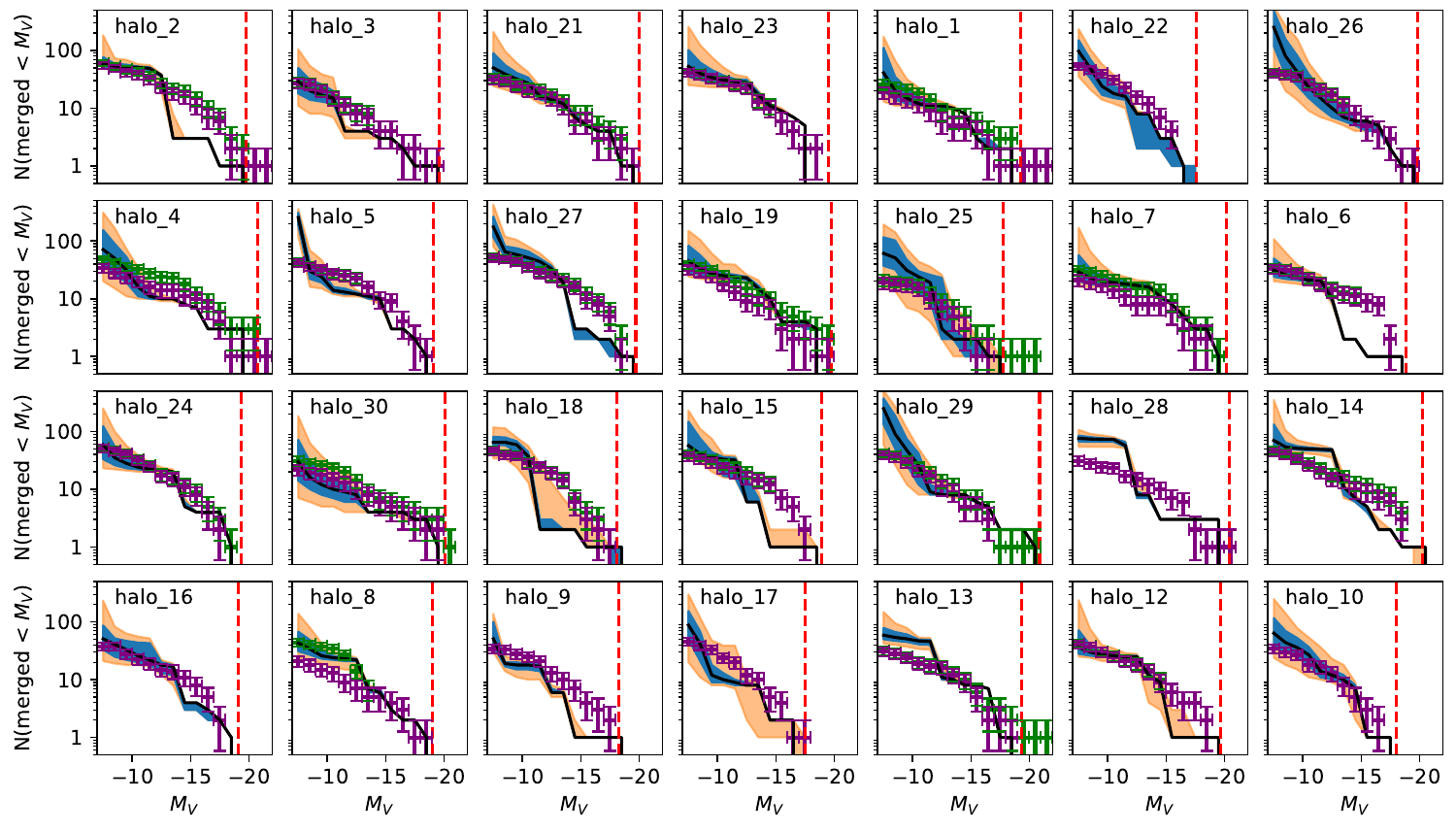}
    \caption{The estimated cumulative number of destroyed dwarfs for the $N=28$ Auriga haloes. The solid black line shows the median, and the shaded blue(orange) regions the 16-84(1-99) percentiles.  The red dashed line indicates the assumed total luminosity of the halo. For each halo, accreted star particles are selected within $r < 20$ kpc. The points with (Poisson) error bars indicate the `truth', with all progenitors shown in green, and only those accreted $> 5$ Gyr ago in purple. The latter are shown because recently accreted dwarfs are likely (i) not fully phase-mixed, and (ii) can significantly deviate from the mass-metallicity relation for (old) dwarf galaxies in Auriga.}
    \label{fig:auriga}
\end{figure*}

\cite{fattahi2020} show using the Auriga simulation suite that the number of destroyed dwarfs in MW-mass haloes is larger than the number of surviving satellites, at least down to $M_V \sim -8$. This is in agreement with our results, however, our estimated \textit{total} number of destroyed dwarfs is far higher than these models (by a factor of $\sim 3-10$, see also Fig. \ref{fig:auriga}). This could be a genuine tension with the models, but it is worth stressing that our number estimates at low luminosities are likely biased high, and the numbers could be reduced if we had larger sample sizes and/or the metal-poor tails of higher mass systems are taken into account (see Section \ref{sec:caveats}).

Finally, given the heterogeneous nature of our sample of halo stars, we consider how different cuts in surface gravity affect the results. Namely, dwarf stars and giants can have different metallicity biases, and probe different volumes in magnitude-limited surveys. The MDF of our halo sample split by log$(g)$ was shown in Fig. \ref{fig:mw_sample}. Here, we can see there are slight differences for low and high log$(g)$, and now we consider how our inferred number counts of destroyed dwarfs are affected. The cumulative number of destroyed dwarfs is shown in Fig. \ref{fig:mw_halo_logg} with two different bins of $\mathrm{log} (g)$, appropriate for dwarf stars ($\mathrm{log} (g) > 3.5$) and giants ($\mathrm{log} (g) <3.5$). It is worth bearing in mind that our overall sample is dominated by the high surface gravity dwarf stars (approximately $\sim 2/3$ have $\mathrm{log} (g) > 3.5$).  Note that here we only use the \cite{kirby2011} mass-metallicity relation, and the total halo luminosity is fixed. Encouragingly, the total number of progenitors (for $M_V \lesssim -10$) is very similar for the two bins of $\mathrm{log} (g)$. However, massive progenitors ($L \gtrsim 10^8L_\odot$) are only favoured in the high $\mathrm{log} (g)$ sample. This is likely because the MDF is biased towards lower metallicities for the giant star sample (see Fig. \ref{fig:mw_sample}). Moreover, the giant and dwarfs are probing slightly different volumes, with the high surface gravity dwarfs more concentrated around the solar neighbourhood. This exercise highlights the difficulty of using a `hodge-podge' of halo stars for our analysis, and it will clearly be preferable for future work to have a more homogeneous sample, where the selection function is clearly defined.

\section{Auriga simulations}
\label{sec:sims}

Our modelling procedure makes various assumptions and simplifications. For example, it assumes each progenitor galaxy is sampled in a representative way, and that their MDFs are adequately described by a Gaussian distribution. In reality, this may not be the case, particularly for volume-limited Galactic-sized stellar haloes. To this end, we test our model on simulated MW stellar haloes, which are representative of `realistic' accreted populations. We apply our modelling procedure to halo stars in the Auriga simulations \citep{grand2017}; these cosmological hydrodynamical simulations are a suite of $N \sim 30$ high resolution ($m_p \sim 5 \times 10^4 M_\odot$) MW-mass ($1-2 \times 10^{12}M_\odot$) haloes. In this work, we make use of the $N=28$ haloes studied in \cite{fattahi2019}, which omits two haloes currently undergoing major mergers. We only consider accreted halo stars, which are identified in \cite{fattahi2019} as those that formed in subhaloes other than the main progenitor galaxy. 

For each halo, we construct a sample of halo star particles within $r <20$ kpc. This is chosen to roughly mimic the volume limit of current observations, and ensure large enough sample sizes. The input into the model is the [Fe/H] values of the stellar particles. Of course, in the simulations, we also know the progenitor galaxy of each star particle, and can thus test the estimated mass spectrum of accreted dwarfs from our modelling procedure. The final ingredient we need to define is the mass-metallicity relation for the Auriga simulations. \cite{grand2021} show that the mass-metallicity relation for dwarf galaxies in Auriga is in good agreement with low-mass dwarfs ($M_{\rm star} \sim 10^6 M_\odot$), but is too metal-rich by $\sim 0.5$ dex for more massive dwarfs (see Figure 13 in \citealt{grand2021}). We use all the destroyed dwarf progenitors across the $N=28$ Auriga haloes to calibrate this relation\footnote{To clarify, all destroyed dwarfs are used, not just those that have debris within 20 kpc of the host halo}. However, we do exclude dwarfs that are accreted recently (less than 5 Gyr ago) as these can have significantly different metallicities due to ongoing star formation. The debris from these events is still included in the analysis, but our calibration is only based on the relatively old dwarf galaxies. Note that we only consider dwarf progenitors with $M_V > -7$, which corresponds to a stellar mass of $M_{\rm star} > 10^5 M_\odot$ or $N > 2$ star particles. We use the `peak' stellar mass of each dwarf, which corresponds to the maximum stellar mass the progenitor has reached. Note that we get similar results if the stellar mass at infall is used instead. The resulting mass-metallicity\footnote{Note that we assume a stellar mass-to-light ratio of $(M/L) =2$ to convert stellar mass to luminosity.} relation for Auriga is: [Fe/H] $= -1.69 + 0.39 \times (\mathrm{log}_{10} L - 6)$. To estimate the scatter around this mean relation, we calculate the scatter for each individual halo, and use the median value across all haloes. This results in a scatter around the mean [Fe/H] relation of 0.3 dex. Finally, we consider the spread in [Fe/H] for individual dwarfs. Unlike the observations, we find no strong evidence for a variation with dwarf mass, so instead adopt a constant dispersion of $0.4$ dex of the MDF for all dwarfs. Armed with the mass-metallicity relation appropriate for Auriga, we can now test our modelling procedure on these cosmological haloes. 

When applying our method to the Auriga haloes, we assume the total luminosity of the halo is known. This of course results in additional uncertainty in the real observations, but we particularly want to investigate the systematic influences present in the cosmological simulations. We consider accreted dwarfs in the range $-7 > M_V > -22$, and estimate the number of dwarfs in 15 bins with bin size of 1 mag. Fig. \ref{fig:auriga} shows the resulting cumulative number of destroyed dwarfs in the Auriga haloes. Each panel shows a different halo, and our estimated numbers are shown with the solid black lines (median), and blue/orange shaded regions (16-84/1-99 percentiles). The points with error bars are the true values, with Poisson noise adopted for the uncertainties in each $M_V$ bin. Note that the `true' values include all dwarfs that have deposited \textit{any} material within 20 kpc of the host halo. Thus, there can be cases where only a small fraction of a destroyed dwarf is included in the sample (see below). The green values in Fig. \ref{fig:auriga} are for \textit{all} progenitors, while the purple are only those accreted earlier than $5$ Gyr ago. In many cases, there is little difference between the green and purple values, because most dwarfs are accreted at earlier times. However, we highlight the most recently accreted dwarfs because these are likely not fully phase-mixed, and can significantly deviate from the mass-metallicity relation for (old) dwarf galaxies in Auriga (see above). In reality, we find that these recently accreted dwarfs only cause a significant effect if the progenitors are relatively massive (e.g. Halo 25). 

We discuss these results more quantitatively below, but first cast a qualitative eye on Fig. \ref{fig:auriga}. In general, our estimates agree well with the true mass spectrum of accreted dwarfs. However, in some cases, there can be notable differences. We find that the most significant deviations are due to the following: (1) relatively massive progenitors that lie off the mass-metallicity relation (e.g. Halo 2, 6) and/or (2) progenitors with a low fraction of their material within the given radial range (e.g. Halo 15, 27). These systematics, and sometimes the combination of both, are most likely to cause our method to fail. On the other hand, there are a significant number of haloes for which we recover the mass spectrum very well, which is encouraging given the complexity of these hydrodynamic simulations, and the cosmological nature of their assembly histories.

\begin{figure}
    \centering
    \includegraphics[width=\linewidth,angle=0]{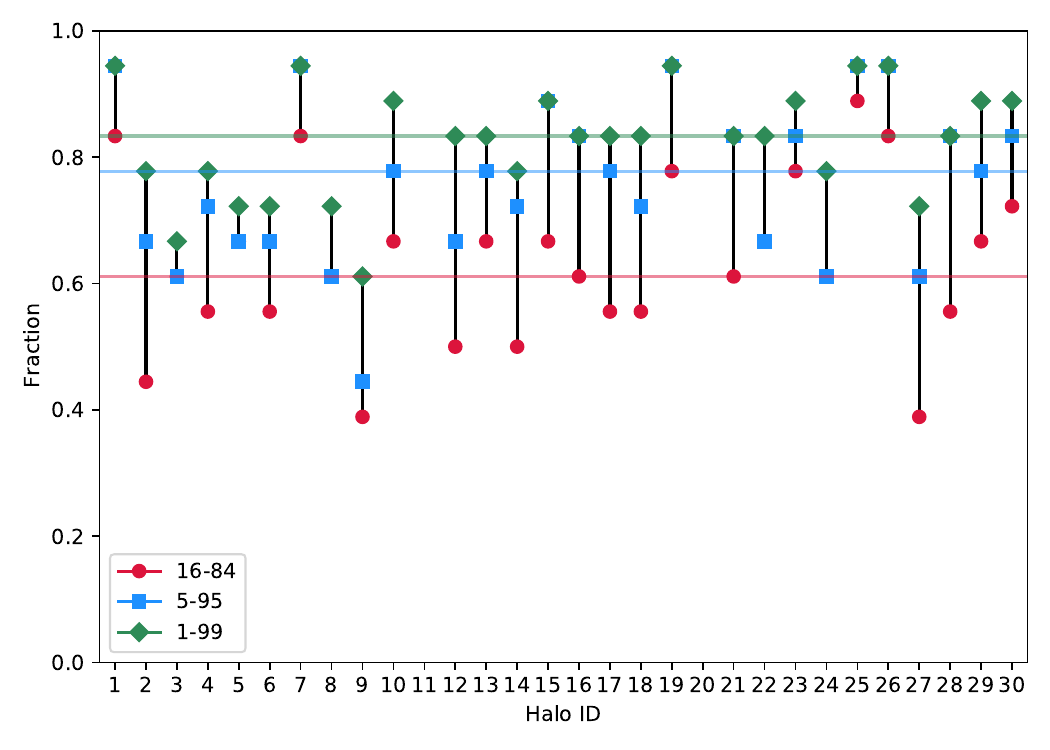}
    \caption{Quantifying the test with Auriga haloes. For each halo, we show the fraction of $M_V$ bins (1 mag wide) that have estimated numbers that agree within the $16-84$ (red-filled circles), $5-95$ (blue-filled squares), and $1-99$ (green-filled diamonds) percentage confidence limits, respectively. The median values are shown with the horizontal coloured lines.}
    \label{fig:auriga_stats}
\end{figure}

\begin{figure*}
  \begin{minipage}{0.49\linewidth}
        \centering
        \includegraphics[width=\textwidth,angle=0]{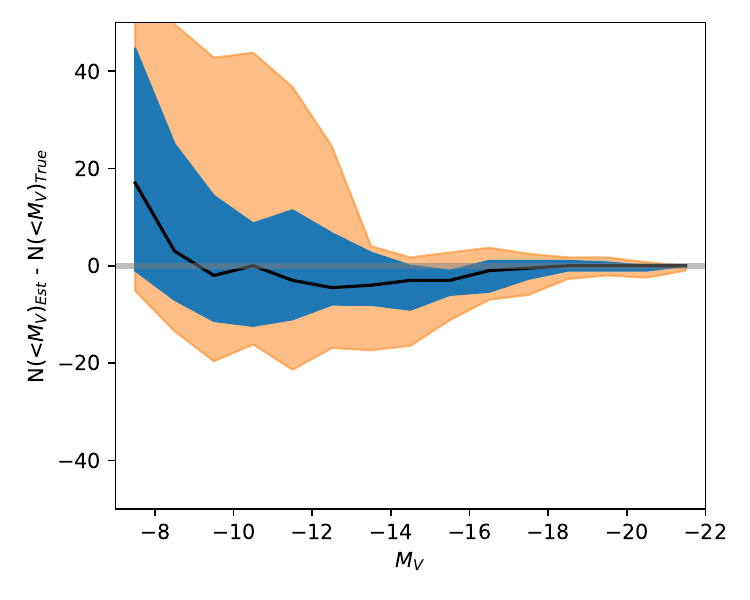}
    \end{minipage}
     \begin{minipage}{0.49\linewidth}
       \centering
           \includegraphics[width=\textwidth,angle=0]{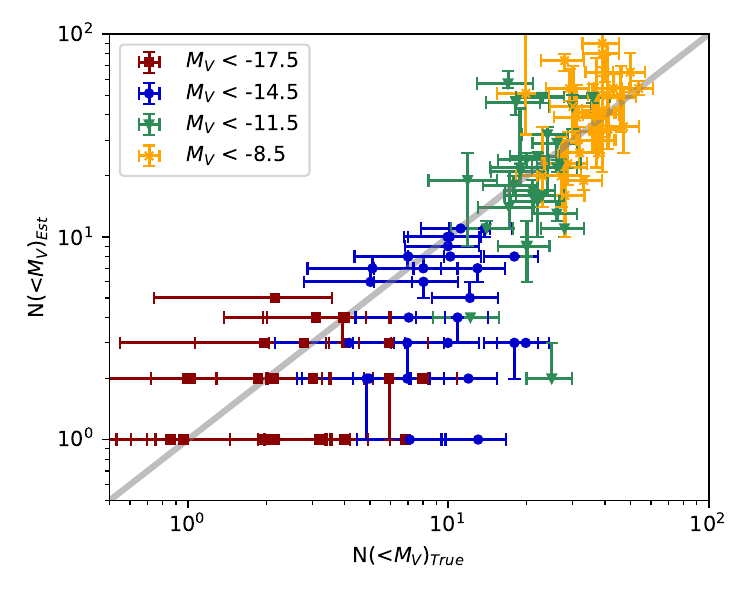}
       \end{minipage}
       \caption{Comparing the estimated and true numbers of destroyed progenitors in the ($N=28$) Auriga haloes. \textit{Left panel:} We show the difference between the estimated and true (cumulative) numbers as a function of $M_{V}$. The solid black line shows the median, and the shaded blue(orange) regions the 16-84(1-99) percentiles. \textit{Right panel:} The estimated versus the true number in different ranges of $M_V$. Error bars show the Poisson errors in the true numbers and the 16-84 percentiles in the estimated numbers. }
       \label{fig:auriga_stats2}
\end{figure*}

In Fig. \ref{fig:auriga_stats} we give a more quantitative summary of our tests of the Auriga haloes. Here, for each halo (identified in the x-axis) we show the fraction of $M_V$ bins that have number estimates that agree within the 16-84, 5-95, and 1-99 percentile confidence limits. The median recovery fractions across all haloes are 0.61, 0.77, and 0.83, respectively. These fractions are below the expected fractions for a `perfect' procedure, but this is unsurprising given the various systematic influences present in the simulations, such as deviations from the adopted mass-metallicity relation and the presence of stellar debris that does not fully occupy the available phase-space. These, of course, are realistic effects that could be present in the observational data.

In Fig. \ref{fig:auriga_stats2} we explore the halo-to-halo scatter more closely. In the left-hand panel, we show the difference between the estimated and true cumulative numbers of destroyed dwarfs as a function of $M_V$. The black line shows the median of the $N=28$ haloes, and the blue and orange shaded regions show the 16-84 and 1-99 percentiles, respectively. The deviation from the true numbers is fairly symmetrical and only starts to shift from zero for very low-mass progenitors. 
It is worth noting that there is a trend toward overestimating the number of accreted dwarfs at lower masses. This could be a real effect, caused by e.g the assumption of Gaussian MDFs, however, this low-mass regime may also be affected by resolution limitations in the simulations, as the MDFs of these dwarfs are only represented by a handful of star particles. In the right-hand panel, we show the estimated vs. true cumulative number of progenitors in four different magnitude ranges. Here, we can see that there is considerable scatter around the 1-to-1 line, but the spread is fairly symmetrical. Finally, to quantify these findings we compute the typical accuracy of our $N(< M_V)$ estimates (averaged over all $M_V$ bins); we find that $N(< M_V)_{\rm est} / N(< M_V)_{\rm true} = 0.9^{+0.6}_{-0.4}$. Thus, we estimate that our method is able to recover the true $N(< M_V)$ within 50\% for most $M_V$ bins.  

\section{Discussion}
\label{sec:disc}

\subsection{Caveats and potential improvements}
\label{sec:caveats}
Probably the most significant caveat in our modelling approach is the assumption of Gaussian MDFs. We know that galaxies are expected to have  metallicity distributions that are not Gaussian \citep{revaz2009,kirby2011,kirby2013}. The details of non-Gaussianity heavily depend on the star formation history, the timescale and intensity of gas inflows \citep{lanfranchi2004,roman2013}, and likely other processes. The non-Gaussianity is likely a bigger problem for more luminous systems, as they have more extended star-formation histories compared to faint systems, which, in some cases, are consistent with a single burst of star formation before reionization \citep{weisz2014}. 
What is the possible systematic effect of neglecting the non-Gaussianity? Assuming that the non-Gaussianity is not caused by accreted systems, but is intrinsic, that would lead us to overestimate the number of accreted fainter systems. Thus our constraints would be upper limits on the number of accreted events. However, the Gaussian assumption is something that can be potentially fixed in our formalism. For example, it could be done by assuming parametric MDF families from \cite{kirby2011}, where one would need to assume some dependence of the MDF parameters on galaxy luminosity.

Another key assumption is that all of the accreted stars are coming from dwarf galaxies. However, it is likely that  some fraction of stars (at least in the MW) are coming from disrupted globular clusters. If trends seen in more massive galaxies extend to faint dwarfs \citep{forbes2018,huang2021,eadie2022}, we may expect that 0.1-1\% of stars come from disrupted GCs. Note, however, that other works have argued for much higher fractions \citep[e.g.][]{martell2011}.  The metallicity distribution of clusters is poorly understood, and since individual GCs have extremely narrow MDFs it is unclear if there is a solution to take GCs into account in our model.

\subsection {Future prospects}
\label{sec:future}

\begin{figure*}
        \centering
        \includegraphics[width=\textwidth,angle=0]{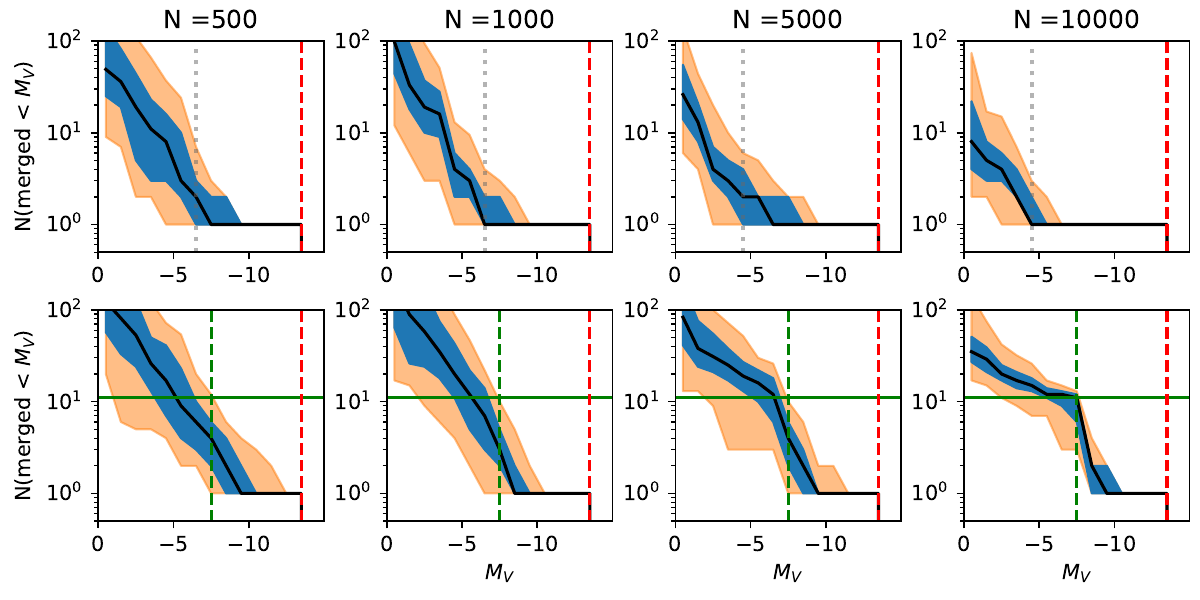}
    \caption{Testing the method on dwarf galaxies with toy fake data. Here, dwarfs are generated with Gaussian MDFs following the adopted $z=0$ mass-metallicity relation. In the top row, there are no merger events (just the MDF of the central galaxy, $M_V=-13.5$). The dotted grey line indicates the approximate $M_V$ value where the estimated number of contributors becomes less reliable. As the sample sizes increase we can probe to fainter luminosity systems. In the bottom row, $N=10$ low mass ($M_V= -7.5$) systems are included. The size of the samples generated increases with each column. The solid black line shows the median, and the shaded blue(orange) regions the 16-84(1-99) percentiles. The vertical red dashed line indicates the $M_V$ of the central galaxy, and the green dashed line shows the $M_V$ of the accreted system (if included). The true number of lower-mass systems is shown with the solid horizontal green line. }
    \label{fig:fake_dwarfs}
\end{figure*}

\begin{figure*}
        \centering
        \includegraphics[width=\textwidth,angle=0]{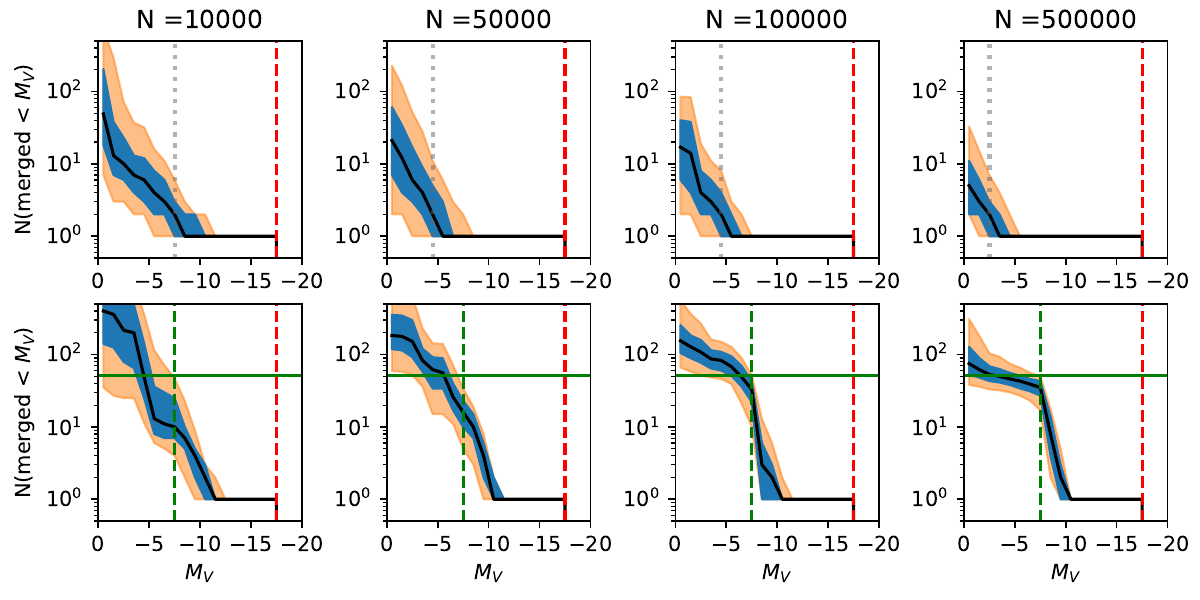}
    \caption{Same as Fig. \ref{fig:fake_dwarfs} but for MW haloes. Here, one massive progenitor is generated ($M_V = -17.5$) with no other progenitors (top panel), and with $N=50$ additional low-mass progenitors (bottom panel).}
    \label{fig:fake_MW}
\end{figure*}

The MDF modelling procedure we have outlined in this work has compelling potential when applied to future spectroscopic data sets. In particular, the availability of much larger numbers of stars with metallicity measurements will allow us to probe to lower dwarf mass scales, and potentially constrain the number of destroyed ultra-faint dwarfs. These latter measurements would not only inform us about the low-mass accretion history of galaxies but could also be used to constrain small-scale galaxy formation and the nature of dark matter \citep{deason2022}.

Here, we use toy models to estimate the sample sizes needed to probe down to the ultra-faint mass scale ($M_V \gtrsim -8$). Note here we focus on the \textit{ideal} case and ignore the potential caveats discussed in the previous sub-section and elsewhere. We generate Gaussian MDFs that follow the \cite{kirby2011} mass-metallicity relation, with varying sample sizes. We consider two example cases, one similar to a classical dwarf galaxy ($M_V= -13.5$), and another akin to a Galactic stellar halo with one main progenitor ($M_V=-17.5$). For each case, we generate the central MDFs with no lower mass progenitors, or with an additional $N=10-50$ low luminosity systems ($M_V=-7.5$). The results of this exercise are shown in Figs. \ref{fig:fake_dwarfs} and \ref{fig:fake_MW}.

It is immediately clear that as the sample sizes increase, we can probe to lower mass scales. The grey dotted line in the top-row of Figs. \ref{fig:fake_dwarfs} and \ref{fig:fake_MW} indicates where the estimated number of contributors starts to become less reliable (i.e. when $N_{\rm Merged} > 1$ for fake tests with no merger events). For the typical sample size of the classical MW satellites ($N \sim 500$) we can currently only reliably probe down to $M_V \gtrsim -7$. For Galactic haloes with $N \sim 10^4$ tracers, we can likely probe down to $M_V \gtrsim -8$.
In order to probe down to the ultra-faint regime ($M_V \gtrsim -5$) requires significant sample sizes that are not currently available. For example, for a typical classical dwarf $N \gtrsim 5000$ stars are needed to unambiguously detect low-mass progenitors. On the other hand, for Galactic stellar haloes the sample sizes likely need to exceed $N \gtrsim  10^5$. Although these numbers are larger than the sample sizes currently available, they are achievable with upcoming spectroscopic surveys. Indeed, the large field-of-view and copious number of fibres available in the DESI, WEAVE, and 4MOST instruments, make them ideal tools for this task. Dedicated programs focusing on classical dwarf satellite galaxies could yield thousands of member stars with spectroscopic measurements. Furthermore, the MW surveys planned with these facilities are predicted to obtain measurements for $N \sim 10^6$ halo stars between $10-30$ kpc \citep[e.g.][]{desi_mw}. These survey data will not only provide significant numbers of dwarf members and halo stars with metallicity measurements, but will also provide more homogeneous sampling, and well-defined selection functions. This latter point is a particular downside of the current implementation in this work, which relies on a combination of data samples with ill-defined selection functions. In summary, the method we propose here is poised to exploit upcoming data sets to robustly quantify the accreted populations of stars in the MW and its dwarf galaxies.

\section{Conclusions}
\label{sec:conc}
We have introduced a new statistical method to model the MDF of a stellar population as an ensemble of individual components. These components follow the galaxy mass-metallicity relation and are assumed to be Gaussian distributed around their mean values (with a mass-dependent spread). We apply the method to observations of the MW halo and classical dwarf satellites, and we also test the procedure on cosmological hydrodynamical simulations of MW-mass haloes. Our main conclusions are summarized as follows:

\begin{itemize}
  
\item Most samples of stars associated with MW dwarf satellites are too small to robustly probe lower mass accretion events. However, we do not find any evidence for significant mergers, and can indeed in some cases (e.g. Fornax, Leo I), rule out accreted components more massive than $M_{V, \rm host} +5$ (or $L_{\rm host}/100$). 

\item We constructed a sample of MW halo stars within $r < 10$ kpc using several spectroscopic surveys and \textit{Gaia} data. By adopting the mass-metallicity relation applicable to surviving dwarf galaxies we find that one massive progenitor is favoured with $L \sim 10^{8.5}L_\odot$, and there are several hundred ($N \sim 400$) progenitors in total down to $M_V < -10$.

\item We also consider a mass-metallicity relation more appropriate for \textit{destroyed} dwarf galaxies for the MW stellar halo, as suggested by \cite{naidu2022}. Here, $N=3$ massive progenitors are favoured, but the total number of progenitors down to $M_V < -10$ is similar to the fiducial case. By placing a stringent constraint on the total halo luminosity ($L_{\rm tot} = 10^9M_\odot$), the two different mass-metallicity relations give more similar results for massive progenitors, but the total number of progenitors differs more significantly (by a factor of 3).

\item We find that the total halo luminosity in our model, and the adopted mass-metallicity relation, are both important assumptions. The former can be constrained by other means \citep[e.g.][]{deason2019, mackereth2020}, and more work needs to be done to understand the redshift evolution of the stellar mass-metallicity relation\footnote{Although the study of the evolution of the mass-metallicity relation is in its infancy, there are several efforts in this direction \citep[e.g.][]{choi2014, leethochawalit2018,leethochawalit2019,beverage2021,zhuang2022}.}

\item Our modelling procedure is applied to the hydrodynamic cosmological Auriga simulations, a suite of $N\sim 30$ MW-mass haloes. Here, many of our assumptions (e.g. phase-mixed material, Gaussian MDFs) are unlikely to hold, so this provides a strong test for our method. We find that, in many cases, our procedure works well, and most failures come from scatter in the mass-metallicity relation and/or recent accretion events not fully occupying the phase-space we are probing. In general, we find that we can recover the true luminosity function ($N(<M_V)$) of destroyed dwarfs to within 50\% for most $M_V$ bins.

\item Finally, we consider how the increase in sample sizes from future spectroscopic surveys can allow us to probe down to the ultra-faint dwarf mass scale ($M_V > -10$). We find that MW stellar halo samples with $N \sim 10^6$ tracers will allow us to probe down to $M_V > -10$; encouragingly, this should be feasible with upcoming surveys such as DESI and WEAVE. Moreover, with sample sizes exceeding $N \sim 5000$ we should be able to probe the lower mass accretion events associated with classical dwarf satellites in the MW. Our ability to probe down to these puny stellar systems will enable us to address fundamental questions about galaxy formation at the lowest mass scales and, potentially, the nature of dark matter.

\end{itemize}

We have shown that using \textit{only} the MDF of an (accreted) stellar population, the mass-spectrum of its progenitors can be uncovered. This is encouraging for the upcoming generation of spectroscopic surveys of the MW. However, a possible extension of this work would be to combine the MDF modelling with phase-space data and/or additional chemical dimensions (see e.g. \citealt{cunningham2022}). The addition of dynamical information could provide tighter constraints on the luminosity function of destroyed dwarfs. In particular, where the MDF modelling is weakest, i.e. when the stellar material is un-mixed in phase-space, is likely where the dynamical data is the most informative. Moving forward, modelling in the chemodynamical space is the next logical step, and, importantly, we will have the data to do this. Thus, it is clear that future data sets combined with modelling methods such as that presented here will provide all the tools needed to finally quantify the accretion history of the Galaxy and its satellite population.

\section*{Acknowledgements}
We thank an anonymous referee for providing useful comments that helped improve the paper.

AD is supported by a Royal Society
University Research Fellowship. AD acknowledges support from the Leverhulme Trust and the Science and Technology Facilities Council (STFC) [grant numbers
ST/P000541/1, ST/T000244/1]. AF is supported by a UKRI Future Leaders Fellowship (grant no MR/T042362/1). RG acknowledges financial support from the Spanish Ministry of Science and Innovation (MICINN) through the Spanish State Research Agency, under the Severo Ochoa Program 2020-2023 (CEX2019-000920-S).

This work used the DiRAC@Durham facility managed by the Institute for Computational Cosmology on behalf of the STFC DiRAC HPC Facility (\url{www.dirac.ac.uk}). The equipment
was funded by BEIS capital funding via STFC capital grants ST/K00042X/1, ST/P002293/1, ST/R002371/1 and ST/S002502/1, Durham University and STFC operations grant ST/R000832/1. DiRAC is part of the National e-Infrastructure. This work was performed in part at Aspen Center for Physics, which is supported by National Science Foundation grant PHY-1607611. This work was partially supported by a grant from the Simons Foundation.

For the purpose of open access, the author has applied a Creative Commons Attribution (CC BY) licence to any Author Accepted Manuscript version arising from this submission. 

AD thanks Ethan Nadler for providing the completeness-corrected estimates of the MW dwarf satellite luminosity function.

\section*{Data Availability}
The data analysed in this article can be made available upon reasonable 
request to the corresponding authors.

The code used to perform the MDF modelling is available on Github\footnote{\url{https://github.com/segasai/mdf_modeling_paper}}



\bibliographystyle{mnras}
\bibliography{references} 




\bsp	
\label{lastpage}
\end{document}